\documentclass{pasj00}
\draft

\newcommand{{\ka}}{K$\alpha$}
\usepackage[dvips]{color} 

\begin{document}
\SetRunningHead{Author(s) in page-head}{Running Head}
\Received{2000/12/31}%{yyyy/mm/dd}
\Accepted{2001/01/01}%{yyyy/mm/dd}

\title{X-ray Spectroscopy of the Mixed Morphology Supernova Remnant
W28 with XMM-Newton}

%%% begin:list of authors
% Do NOT capitalize all letters in "textsc".
\author{Ryoko \textsc{Nakamura}$^1$,
Aya \textsc{Bamba}$^2$,
Manabu \textsc{Ishida}$^3$,
Ryo \textsc{Yamazaki}$^3$,
Ken'ichi \textsc{Tatematsu}$^4$,
Kazunori \textsc{Kohri}$^{5,6}$,
Gerd \textsc{P\"{u}hlhofer}$^7$,
Stefan J. \textsc{Wagner}$^8$,
and
Makoto \textsc{Sawada}$^2$
}
\affil{$^1$
Tsukuba Space Center/JAXA,
1-1 Sengen 2chome, Tsukuba-shi, Ibaraki 305-8505, Japan}
\affil{$^2$
Department of Physics and Mathematics, Aoyama Gakuin University
5-10-1 Fuchinobe Chuo-ku, Sagamihara, Kanagawa 252-5258, Japan}
\affil{$^3$
   Institute of Space and Astronautical Science/JAXA,
   3--1--1 Yoshinodai,Sagamihara, Kanagawa 229--8510, Japan}
\affil{$^4$
   National Astronomical Observatory,
   2-21-1 Osawa, Mitaka, Tokyo 181-8588, Japan}
\affil{$^5$
   Theory Center, Institute of Particle and Nuclear Studies, KEK,
   1-1 Oho, Tsukuba 305-0801, Japan}
\affil{$^6$
   The Graduate University for Advanced Studies (Sokendai),
   1-1 Oho, Tsukuba 305-0801, Japan}
\affil{$^7$
   Institut f\"{u}r Astronomie und Astrophysik,
   Sand 1, 72076 T\"{u}bingen, Germany}
\affil{$^8$
Landessternwarte, Universit\"{a}t Heidelberg,
K\"{o}nigstuhl, 69117 Heidelberg, Germany}
\email{sawada@phys.aoyama.ac.jp}
% \altaffiltext{1}{Address of Institute}
% \email{ddddd@xxx.xxx.xx.xx}
% \email{eeeee@xxx.xxx.xx.xx}
% \altaffiltext{2}{Address of Institute}

%% `\KeyWords{}' always has to be placed before `\maketitle'.
 %Do NOT move this preamble from here!

\maketitle

\begin{abstract}

 We report on spatially resolved X-ray spectroscopy of the north-eastern
 part of the mixed morphology supernova remnant (SNR) W28 with {\it
 XMM-Newton}. The observed field of view includes a prominent and
 twisted shell emission forming the edge of this SNR as well as part of
 the center-filled X-ray emission brightening toward the south-west edge
 of the field of view.  The shell region spectra are in general
 represented by an optically thin thermal plasma emission in collisional ionization
 equilibrium with a temperature of $\sim$0.3~keV and
a density of $\sim$10~cm$^{-3}$,
which is much higher than the density obtained for inner parts.
% In the inner region spectra, on the
% other hand, we discovered a power-law emission with a photon index of
% $\sim$3, which is probably a non-thermal emission caused by particle
% acceleration. A similar temperature thermal emission was also detected
% from the inner region, although it is in collisional non-equilibrium
% ionization state with an ionization time scale of $\simeq$10$^{11}$
% cm$^{-3}$s. The density the ionization age of the plasma
% were obtained to be $\sim$0.5~cm$^{-3}$ and several times $10^4$~yr,
% respectively.
In contrast, we detected no significant X-ray flux
 from one of the TeV $\gamma$-ray peaks with an upper-limit flux of
 2.1$\times$10$^{-14}$ erg cm$^{-2}$ s$^{-1}$ in the 2--10~keV band. The
 large flux ratio of TeV to X-ray, larger than 16,
and the spatial coincidence of the
 molecular cloud and the TeV $\gamma$-ray emission site indicate that
 the TeV $\gamma$-ray of W28 is $\pi^{0}$-decay emission originating
 from collisions between accelerated protons and molecular cloud
 protons. Comparing the spectrum in the TeV band and the X-ray upper
 limit, we obtained a weak upper limit on the magnetic field strength
 $B\lesssim$ 1500~$\mu$G.

\end{abstract}

\section{Introduction}

Supernova remnants (SNRs) are one of the most promising acceleration
sites of cosmic rays up to $\sim10^{15.5}$~eV~(the knee
energy). \citet{1995Natur.378..255K} discovered synchrotron X--rays from
the shell of SN 1006, indicating the existence of extremely high-energy
electrons up to $\sim$~TeV produced by the first-order Fermi
acceleration. Following this discovery, the synchrotron X-ray emission
has been discovered from a few more young shell-type SNRs, such as
RX~J1713.7--3946 \citep{1997PASJ...49L...7K}, RCW~86
\citep{2000PASJ...52.1157B,2001ApJ...550..334B},
and RX~J0852.0$-$4622 \citep{2000PASJ...52..887T}.
On the other hand, TeV $\gamma$--rays have also been detected from some
SNRs
\citep{2004Natur.432...75A,2006A&A...448L..43A}.
The radiation of TeV $\gamma$-ray is explained by (1)
Inverse-Compton scattering (IC) of cosmic microwave background photons
by the same high energy electron giving rise to the X-ray synchrotron
emission, (2) non-thermal bremsstrahlung by high energy elections or (3)
the decay of neutral pions that are generated by collisions between high
energy protons and dense interstellar matter. 

Most of the SNRs with such a X-ray and TeV $\gamma$-ray evidence
have been young with an age of less than several thousands of years.
%Most of the SNRs from which evidence of cosmic ray acceleration has
%been found are young with an age of less than several thousands of years,
%except
%for the middle-aged SNR G156.2+5.7
%\citep{1999PASJ...51...13Y,2009PASJ...61S.155K}. This fact is very
%interesting, because the current standard theory of diffusive shock
%acceleration requires a high shock velocity of at least $\sim$2000
%km~s$^{-1}$ to accelerate X-ray emitting electrons,
%which is not expected in middle-aged SNRs~\citep{2003A&A...400..567U}.
%\citet{2000ApJ...540..292E} argue that
%the non-thermal X-ray emission stands out
%from the thermal X-ray with a weak ambient magnetic field and/or a low
%ambient density.
Recently, Fermi discovered GeV $\gamma$-rays
from several SNRs,
such as W44 \citep{2010Sci...327.1103A},
IC443 \citep{2010ApJ...712..459A},
W51C \citep{2009ApJ...706L...1A},
and so on.
Interesting fact is that
they are not only young but rather old SNRs
compared with the TeV $\gamma$-ray emitting SNRs.
It can be due to the escape of high energy particles
from the shocks
\citep{2011MNRAS.410.1577O,2011ApJ...731...87E,%
2012MNRAS.421..935L,2012A&A...541A.153T,2013MNRAS.429.1643N}.
Most SNRs with GeV $\gamma$-rays are interacting with molecular clouds,
so it may be related with the escape.
However, it is still unknown the detailed picture 
how particles escapes from the shocks.

%In addition, \citet{2006MNRAS.371.1975Y} calculated the
%TeV $\gamma$-ray emission from old SNRs. In this case, the TeV
%$\gamma$-rays come from the $\pi^{0}$-decay, while the secondary
%electrons arising from charged pion decay may emit the synchrotron
%radiation in the X-ray band.

The SNR W28, locating at ($l$, $b$) = ($\timeform{6D.4}$,
$-\timeform{0D.1}$), is an interesting target as a cosmic-ray
accelerator from which both GeV and TeV $\gamma$-rays were detected from the
eastern edge of the radio
shell~(\cite{2009ApJS..183...46A,2008A&A...481..401A}). The diameter and
the distance to W28 are 48 arcmin and 1.9~kpc,
respectively~\citep{2002AJ....124.2145V}. The age seems to be several
times 10$^{4}$ year, which means the remnant is middle-aged~\citep{2002ApJ...575..201R}. 
W28 is classified as ``mixed--morphology'' SNR, showing center-filled X-rays and a
shell-like radio emission. The shell-like radio emission peaks at the
northern and northeastern boundaries where interaction of the SNR matter
with the molecular cloud is established~\citep{1981ApJ...245..105W}.
This interaction was revealed by a lot of OH maser spots
\citep{1994ApJ...424L.111F,1997ApJ...489..143C,2005ApJ...620..257H},
which are signposts of molecular interactions and the location of high
density shocked gas ($n \ge$10$^{4}$ cm$^{-3}$;
\cite{1999PASJ...51L...7A}). TeV $\gamma$-ray emission was detected
by H.E.S.S.~\citep{2008A&A...481..401A} from the eastern edge of the
radio shell.
Recently, {\it{Fermi}} and {\it AGILE}
also detected $\gamma$-rays at
100~MeV $\textless E \textless$ 100~GeV
~\citep{2010ApJ...718..348A,2010A&A...516L..11G}
from the same region of the TeV emission.
In contrast, the emission in the X-ray band
had been believed to comprise of thermal
radiation. \citet{2002ApJ...575..201R} evaluated temperatures of 1.5~keV
in the southwest, 0.56~keV in the northeast, and the central region
requires two-temperature plasma with 0.6~keV and 1.8~keV. The long
ionization timescales in the northeast and central region imply that the
gas is close to the ionization equilibrium.
Recently, Suzaku discovered that
the thermal X-ray emission from the inner region is over-ionized,
implying that the plasma may underwent the sudden rarefaction
\citep{2012PASJ...64...81S}.
These facts imply that
W28 is an ideal target to study the particle escape from the shock.

We analyzed XMM-Newton archival data of the north--eastern part of
W28, where the molecular clouds, OH maser spots, bright X-ray shells,
GeV and TeV emission were detected. XMM-Newton has large effective
area and high angular resolution. These characteristics enable us to
carry out high quality spatially resolved spectroscopy. In \S~2, we
present the observation log and data reduction method. Imaging and
spectral analyses are shown in \S~3 and \S~4, respectively.
%We have detected non-%thermal power-law component for the first time from
%the inner part of W28, in addition to the thermal component so far
%known.
From one of the TeV $\gamma$-ray emission region,
we obtained only the upper limit on the X-ray flux. Discussions
are made on the basis of these results in \S~5 on the nature of the
thermal and non-thermal components in multi-wavelength. Finally we
summarize our results in \S~6.

%\noindent IMPORTANT NOTICE\\
%1. ``\verb|\draft|'' creates single column and double spaces format.\\
%2. If you comment out ``\verb|\draft|'', the output will be double column
%   and single space.\\
%3. For cross-references, the use of ``\verb|\label|, \verb|\ref|, \verb|\cite|'' 
%   and the thebibliography environment is strongly recommended. \\
%4. Do NOT use ``\verb|\def|, \verb|\renewcommand|''.\\
%5. Do NOT redifine commands provided by PASJ00.cls.\\

%\newpage

\section{Observation and Data Reduction}

The north-eastern part of W28 was observed with the European Photon
Imaging Camera (EPIC) on board XMM-Newton Observatory on 2002
September 23 (ObsID = 0145970101) and 2003 October 7 (ObsID =
0145970401). The nominal pointing position was $\alpha =
-\timeform{270D.438}$, $\delta = -\timeform{23D.300}$ (J2000). All of
the EPIC instruments were operated in the full-frame mode with a thick
filter. We used version 7.0.0 of the Standard Analysis System (SAS)
software, and selected X-ray event with PATTERN keywords of $\le$12 for
the MOS1/2 and $\le$4 for the pn, respectively.

The net exposure times were 54.1~ks and 49.9~ks for the MOS1/2 and
the pn, respectively, after combining the 2002 and 2003 data. To remove
the high particle background time intervals, we accumulated a lightcurve
in the 10--12~keV band from the whole field of view, and filtered the
time intervals when the count rate was larger than 0.35~counts~s$^{-1}$ for the
MOS and 0.4~counts~s$^{-1}$ for the pn. After this screening, the effective
exposure time of MOS1, MOS2, and pn were 51.9~ks, 52.1~ks and
39.6~ks, respectively.

\section{Image Analysis}

\begin{figure}[htpb]
  \begin{center}
    \includegraphics[width=80mm]{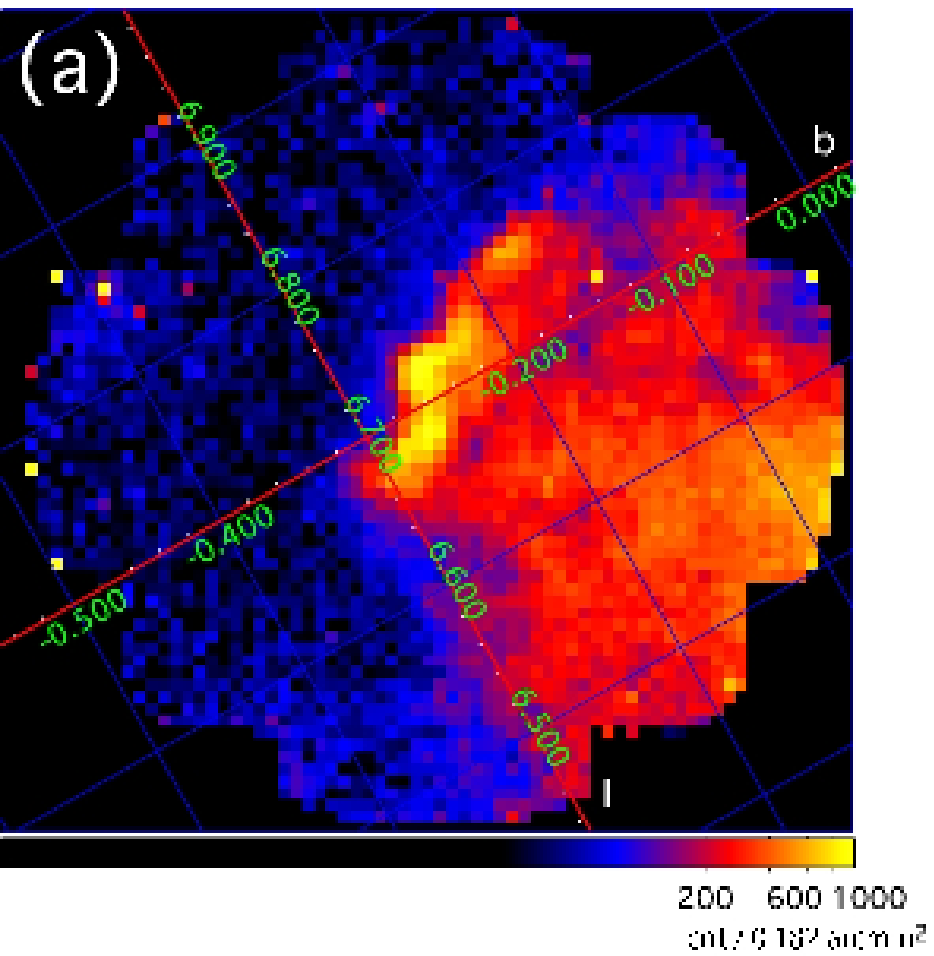}
    \includegraphics[width=80mm]{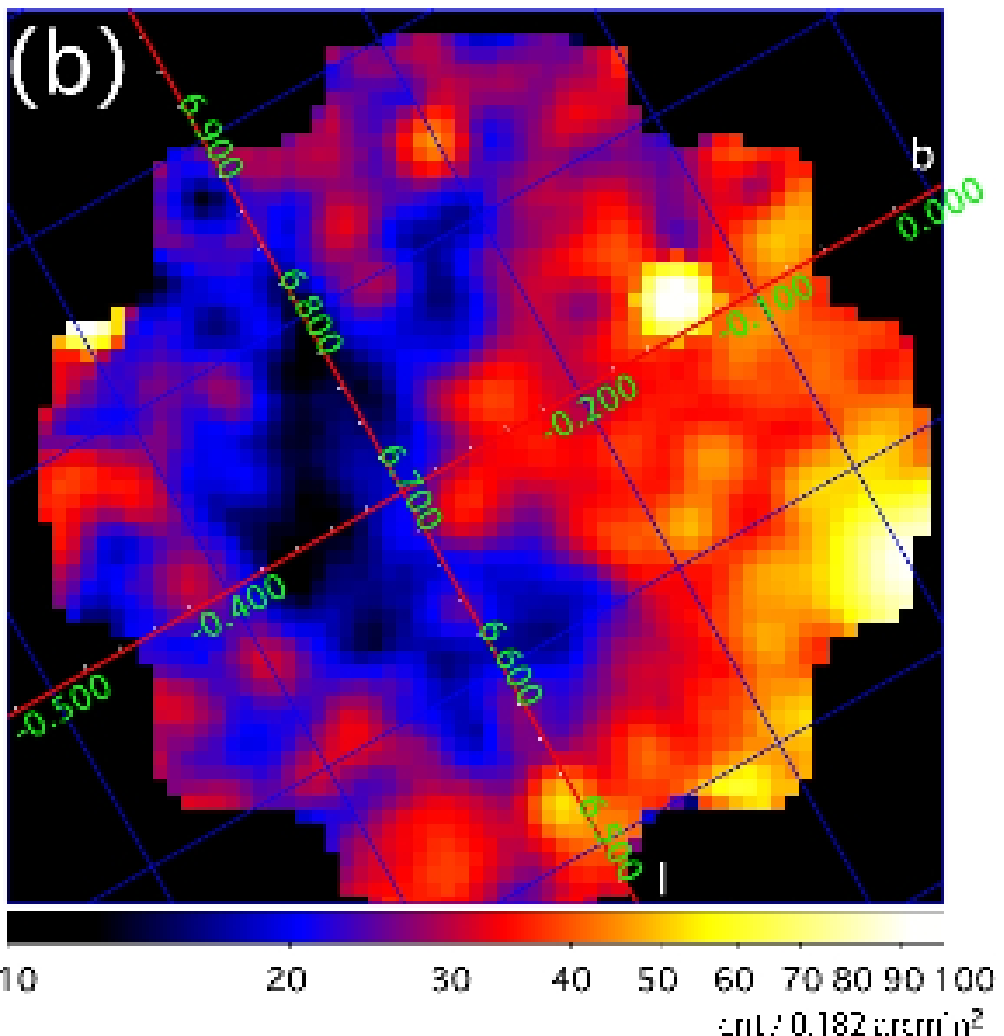}
  \end{center}
  \caption{Exposure corrected images of the north-eastern part of W28
  (a) in 0.3--2.0~keV and (b) in 2.0--10.0~keV in logarithmic scale in
  the galactic coordinates. The images are binned up to
  \timeform{25.''6} per pixel.}\label{fig:ximage}
\end{figure}

Figure \ref{fig:ximage} shows the exposure-corrected MOS images in
0.3--2.0~keV and 2.0--10.0~keV with a binning size of
\timeform{25''.6}. 
They were created by combining all the MOS1/2 data from the 2002 and
2003 observations. In the low-energy band (figure~\ref{fig:ximage}a), the shell region
located at ($l$, $b$) $\simeq$ ($\timeform{6D.70}$, $-\timeform{0D.26}$)
is the brightest. Its shape is twisted in a complex manner. The inner
region of W28, toward the southwest of the image, is also enhanced in
surface brightness. In the high energy band (figure~\ref{fig:ximage}b), on the other hand,
the shell region is much fainter than the inner region.

\begin{figure}[htpb]
  \begin{center}
    \includegraphics[width=80mm]{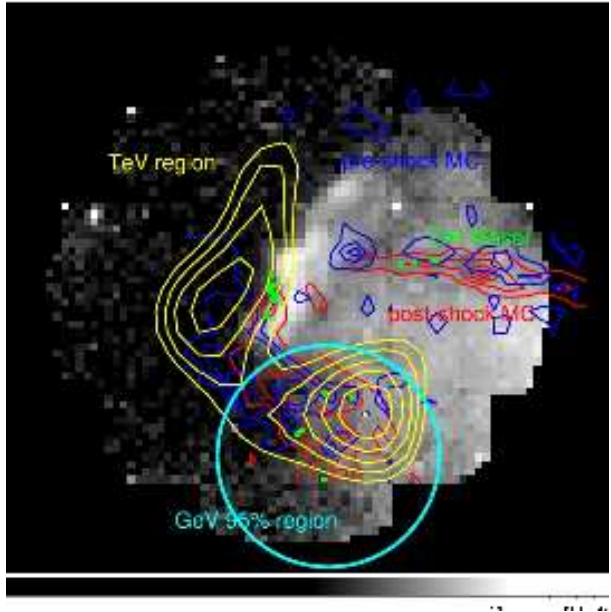}
    %%% \FigureFile(width,height){filename}
  \end{center}
  \caption{Gray scale X-ray image with molecular clouds (the blue and
 red) and TeV $\gamma$-ray (yellow) contours overlaid. The blue and red
 molecular cloud contours are velocity-integrated line intensities of CO (J = 1--0) and CO
 (J = 3--2) in logarithmic scale, respectively, which correspond to
 0--55~K~km~s$^{-1}$ pre-shock cloud and 60--650~K~km~s$^{-1}$ post-shock cloud,
 respectively. The yellow TeV $\gamma$-ray contours are in linear scale
 from 50 to 70~count/1.4~arcmin$^{2}$. Green points are the positions of
 OH maser.
%The 95\% confidence region of a GeV source detected with the
% Fermi satellite is shown in cyan circle.
}\label{fig:allimage}
\end{figure}

In figure~\ref{fig:allimage} shown are intensity contours of CO (J = 3--2)
and CO (J = 1--0) in red and blue \citep{1999PASJ...51L...7A} and those
of TeV $\gamma$-ray measured with H.E.S.S. in yellow \citep{2008A&A...481..401A},
overlaid on the 0.3--2.0~keV gray scale MOS image
(figure~\ref{fig:ximage}a).
The CO J =3--2 and 1--0 contours trace distributions of post-shock and
pre-shock molecular clouds, respectively
\citep{1999PASJ...51L...7A}. The green dots represent OH maser spots,
which also indicate presence of shock waves
\citep{1997ApJ...489..143C}.
%The cyan circle indicates the 95\%
%confident error circle of the GeV $\gamma$-ray source detected with {\it
%Fermi} in \citet{2009ApJS..183...46A}.
GeV emission is from the same region of TeV emission
\citep{2010ApJ...718..348A,2010A&A...516L..11G}.
The eastern bunch of the OH maser
sources spatially coincides with the edge of the X-ray bright shell, as
well as the edge of the eastern molecular cloud. This indicates that the
shock is formed there. Edge of one of the TeV $\gamma$-ray peaks seems to
appear at the same position. OH maser spots are also detected with
spatial coincidence with a molecular cloud region which extends linearly
from the X-ray shell toward western edge of the image. In the southern
part of the image, the molecular clouds coincide with the other TeV
$\gamma$-ray emission peak, where the surface brightness of the X-ray
emission is somewhat reduced. From this region, several OH maser spots
are also detected. They are all within the error circle of the GeV
$\gamma$-ray source.

\section{Spectral Analysis}

In this section, we present the results of spatially-resolved spectral
analysis. We concentrate our analysis on
the shell regions and the eastern peak of the TeV $\gamma$-ray emission,
since the inner region has been studied with good statistics of Suzaku
\citep{2012PASJ...64...81S}.
In evaluating the spectra, we
have utilized {\sc xspec} (version 11.3.2) in the band 0.3--10.0~keV. We
have created ancillary response files (ARF) by assuming flat brightness
distribution within each source integration region. As emission spectral
models, we basically adopt the {\sc nei} model to represent optically
thin thermal spectra in ionization non-equilibrium. In some cases where
the {\sc nei} model indicates the ionization equilibrium, we utilize the
{\sc apec} model also. In applying these models to the data, we adopt
the metal composition of \citet{1989GeCoA..53..197A} as the solar
abundance.  To represent interstellar absorption, we multiply the model
{\sc phabs} on these emission models. In the course of the spectral
fitting, we have found that there remain wiggles in the fit
residual. This is caused by difference of gain among the CCD chips of
MOS-1/2 and the inaccuracy of calibration of the line spread
function. Accordingly, we always multiply a Gaussian smoothing model
({\sc gsmooth} in {\sc xspec}) on the emission models. The errors quoted
are always at the 90\% confidence level.

\subsection{North-eastern shell region}

\begin{figure}[htpb]
  \begin{center}
    \includegraphics[width=70mm]{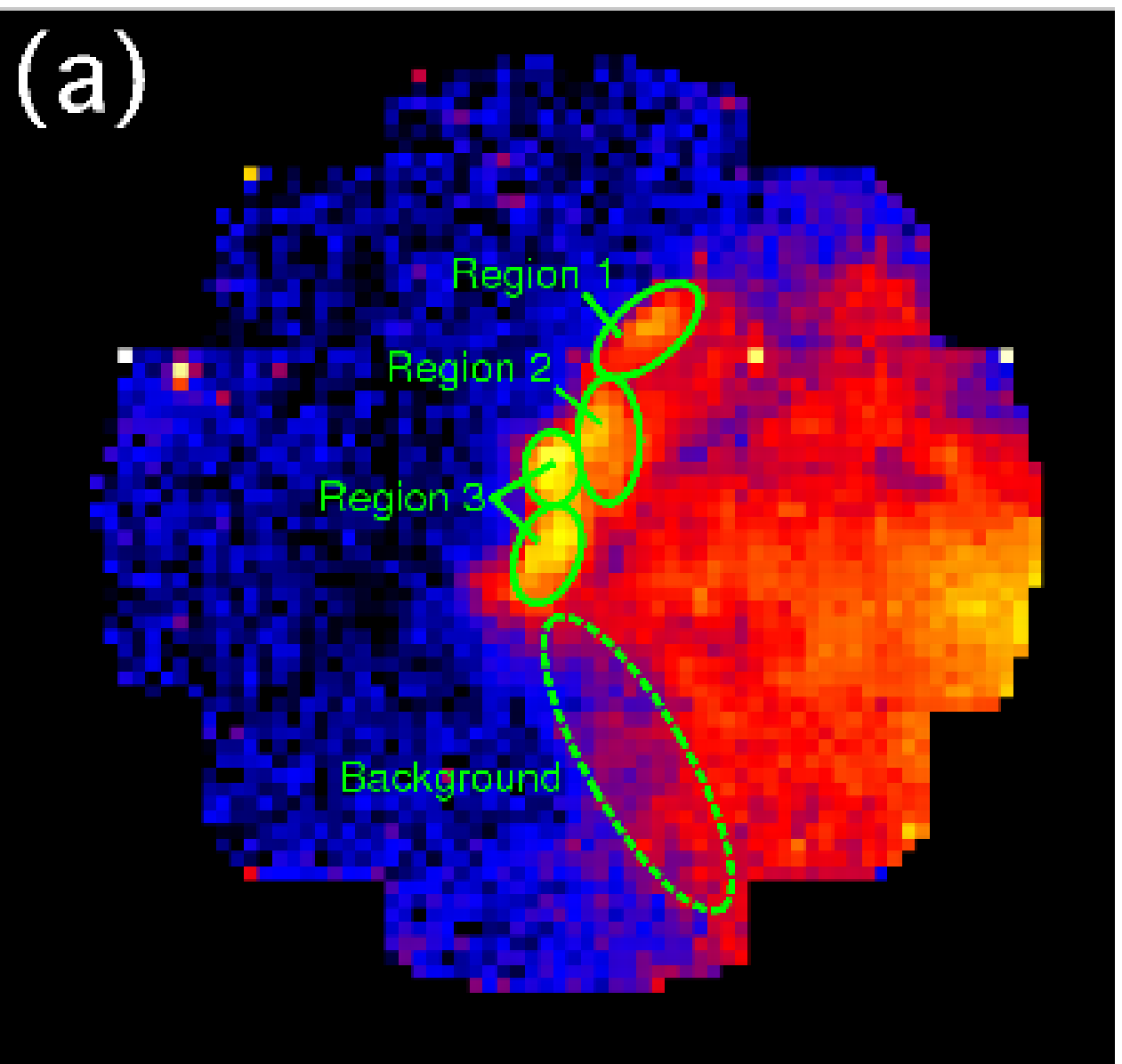}
    \includegraphics[width=90mm]{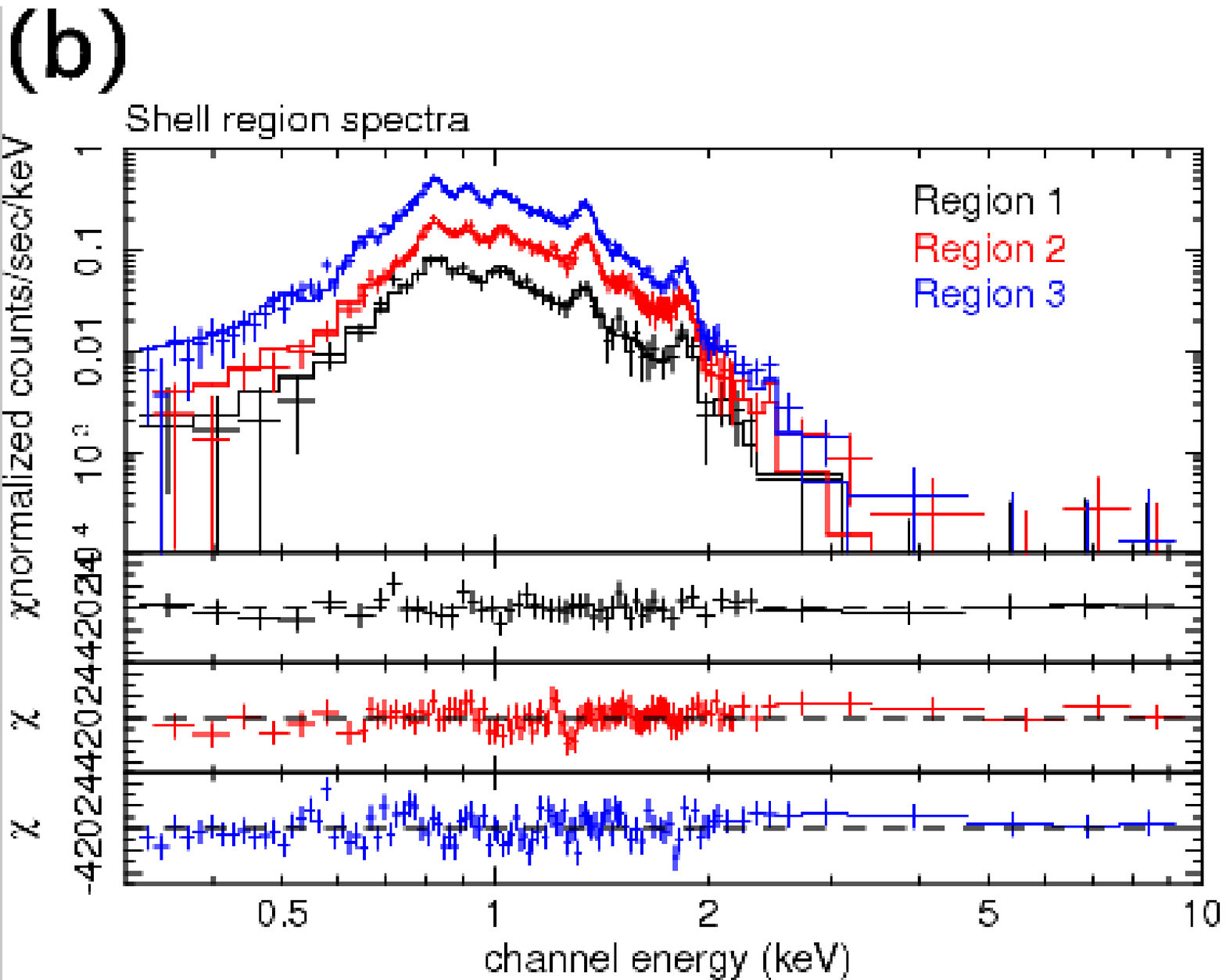}
  \end{center}
  \caption{(a) The photon integration regions for the north-eastern
  shell. Region~1 and 2 are an ellipse with $\timeform{0.'96}$ $\times$
  $\timeform{1.'9}$, and region~3 is composed of two ellipses, one with
  $\timeform{0.'85}$ $\times$ $\timeform{1.'1}$ and the other with
  $\timeform{0.'96}$ $\times$ $\timeform{1.'5}$. The dashed ellipse is
  the common background region. (b)~MOS spectra of each regions with the
  best-fit {\sc vapec} model. Black, red and green represent the data
  and the models of region 1, 2, and 3, respectively. The residuals are
  shown in the bottom panels. The best-fit parameters are summarized in
  table~\ref{tab:param_shell}.}\label{fig:shell}
\end{figure}

% We tried to bring out the feature of bright excess located at the
% northeast shell.
In figure~\ref{fig:shell}~(a) shown with ellipses are regions 1 through 3
for collecting photons from the north-eastern shell, overlaid on the
0.3--10~keV MOS image.
% spectral analysis. We defined the source regions in the order from the
% north, region 1 to 3 shown in figure \ref{fig:shell}~(a) in solid
% ellipses.
% The sizes with a major and minor axis radius are $\timeform{1.'9}$
% $\times$ $\timeform{0.'96}$ for region 1 and 2, $\timeform{1.'1}$
% $\times$ $\timeform{0.'85}$ + $\timeform{1.'5}$ $\times$
% $\timeform{0.'96}$ for region 3, respectively.
%
The dashed ellipse is the background region. This region is taken
symmetrical to regions 1 through 3 with respect to the peak of the
center-filled brightness distribution of W28 as well as to the optical
axis position of the current field of view, in order to elucidate the
nature of an excess emission from the shell.
% region to remove the steady shell emission and to take into account the
% telescope vignetting.

The background-subtracted MOS spectra are shown in figure
\ref{fig:shell}~(b) which are combined ones of MOS1 and MOS2 from the
2002 and 2003 observations.  The black, red and blue crosses represent
the data points from region 1, 2 and 3, respectively. We have 
detected obviously He-like K$\alpha$ emission lines from O (0.57~keV), Ne
(0.91~keV), Mg (1.34~keV), Si (1.86~keV), and L lines of Fe around
1~keV. This means that the spectra include an optically thin thermal
component. We therefore tried to fit the spectra with a single
temperature non-equilibrium collisional ionization plasma emission model
({\sc vnei} model in {\sc xspec};
\cite{2001ApJ...548..820B,1983ApJS...51..115H,1994ApJ...429..710B,1995ApJ...438L.115L})
undergoing photoelectric absorption represented with a single hydrogen
column density $N_{\rm H}$. In the fitting, we set abundances of O, Ne,
Mg, Si, and Fe free to vary but constrained to be common among the
regions, because no statistically significant difference in the
abundances are found among the regions in a trial fit. The abundances of
the other elements are fixed at the solar values. The fitting is
acceptable with the reduced $\chi^{2}$ of 0.91. As a result, however, we
found an ionization parameter $n_{\rm e}t$ of $\sim 10^{13}$~cm$^{-3}$s,
which indicates the shell plasma is in collisional ionization
equilibrium. Accordingly, we replaced the {\sc vnei} model with a {\sc
vapec} model which represents a spectrum from a plasma in collisional
ionization equilibrium\footnote{http://hea-www.harvard.edu/APEC}. The best-fit
parameters are summarized in table~\ref{tab:param_shell}, and the
best-fit models as well as the residuals are displayed in
figure~\ref{fig:shell}~(b).
\begin{table}[htpb]
\begin{center}
\caption{Best-fit parameters of the north-eastern shell region spectra}
\label{tab:param_shell}
 \begin{tabular}{lccc}
 \hline \hline
 \multicolumn{1}{c}{Parameters}       & Region~1 & Region~2 & Region~3 \\
 \hline
 {\sc vapec} & & &\\
  \hspace{1pc}Temperature \ [keV] &  0.37$^{+0.05}_{-0.03}$ & 0.30$^{+0.02}_{-0.01}$ & 0.28~$\pm0.01$\\
  \hspace{1pc}Abundance \footnotemark[a] \ O & \multicolumn{3}{c}{0.29$^{+0.07}_{-0.05}$}\\
   \hspace{6.1pc}Ne &  \multicolumn{3}{c}{0.33$^{+0.06}_{-0.04}$}\\
   \hspace{6.1pc}Mg &  \multicolumn{3}{c}{0.39$^{+0.06}_{-0.05}$}\\
   \hspace{6.1pc}Si &  \multicolumn{3}{c}{0.62$^{+0.10}_{-0.08}$}\\
   \hspace{6.1pc}Fe &  \multicolumn{3}{c}{0.42$^{+0.07}_{-0.05}$}\\
 \hspace{1pc}{\it E.M.} \footnotemark[b] &  1.8$^{+0.6}_{-0.5}$ & 10~$\pm2$ & 22~$\pm3$\\
 $N_{\rm H}$ \footnotemark[c] & 6.2~$\pm0.5$ & 8.2~$\pm0.3$ & 7.5~$\pm0.2$\\ 
 gsmooth & & &\\
\hspace{1pc}$\sigma$~(MOS) \footnotemark[d] & 0.11$^{+0.06}_{-0.05}$ & 0.088$^{+0.030}_{-0.033}$ & 0.094$^{+0.020}_{-0.018}$ \\
\hspace{1pc}$\sigma$~(pn) \footnotemark[d] & 0.11$^{+0.09}_{-0.11}$ & 0.23~$\pm0.05$ & 0.17~$\pm0.03$\\
\hspace{1pc}index \footnotemark[e] & 1.0~(fix) & 1.0~(fix) & 1.0~(fix)\\
$\chi^{2}$/d.o.f~(reduced $\chi^{2}$) & \multicolumn{3}{c}{461.6 / 546~(0.85)}\\
\hline \hline
 \multicolumn{4}{@{}l@{}}{\hbox to 0pt{\parbox{100mm}{\footnotesize
     \footnotemark[a] Abundance ratio relative to the
 solar value (Anders \& Grevesse, 1989). The abundances are common over the regions.\\
    \footnotemark[b] Emission measure {\it E.M.} =
 $\int n_{e}n_{H}dV \simeq n_{e}^{2}V$ in units of 10$^{56}$ \
 cm$^{-3}$, where $ n_{e}$ and $V$ are the electron density and the
  plasma volume. The distance to W28 is assumed to be 1.9~kpc~\citep{2002AJ....124.2145V}. \\
    \footnotemark[c] Absorption hydrogen column density in units of
 10$^{21}$ \ cm$^{-2}$.\\
    \footnotemark[d] Gaussian sigma at 6~keV in a unit of keV.\\
    \footnotemark[e] Energy index of $\sigma$, namely, $\sigma \propto
  E^{\rm -index}$.\\
     }\hss}}
 \end{tabular}
\end{center}
\end{table}
Note that we used both MOS and pn data for spectral fitting, although
only MOS spectra are shown in figure \ref{fig:shell}~(b) for clarity.
The reduced $\chi^2$ of 0.85 implies that the fit is acceptable at the
90\% confidence level. The temperatures are obtained to be
$kT=0.37^{+0.05}_{-0.03}$~keV, $0.30^{+0.02}_{-0.01}$~keV, and
$0.28\pm{0.01}$~keV for regions 1, 2, and 3, respectively. The
temperature decreases from north to south in the shell. 
We note that the temperature we obtained are lower than
that of $\sim$0.6~keV in \citet{2002ApJ...575..201R}.
The temperature difference should be due to the different background regions.
The spectral analysis of background region we used is shown in \S4.2.
The hydrogen
column density $N_{\rm H}$ are (6.2$\pm0.5) \times10^{21}$~cm$^{-2}$,
(8.2$\pm0.3) \times10^{21}$~cm$^{-2}$ and
(7.5$\pm0.2) \times10^{21}$~cm$^{-2}$ for regions 1, 2, and 3,
respectively. The hydrogen column densities of region 2 and region 3
near the molecular cloud are higher than that of region 1.
The abundances of O, Ne, Mg, Si, and Fe
are lower than the solar values.

\subsection{TeV $\gamma$-ray region}
\begin{figure}
  \begin{center}
    \includegraphics[width=40mm]{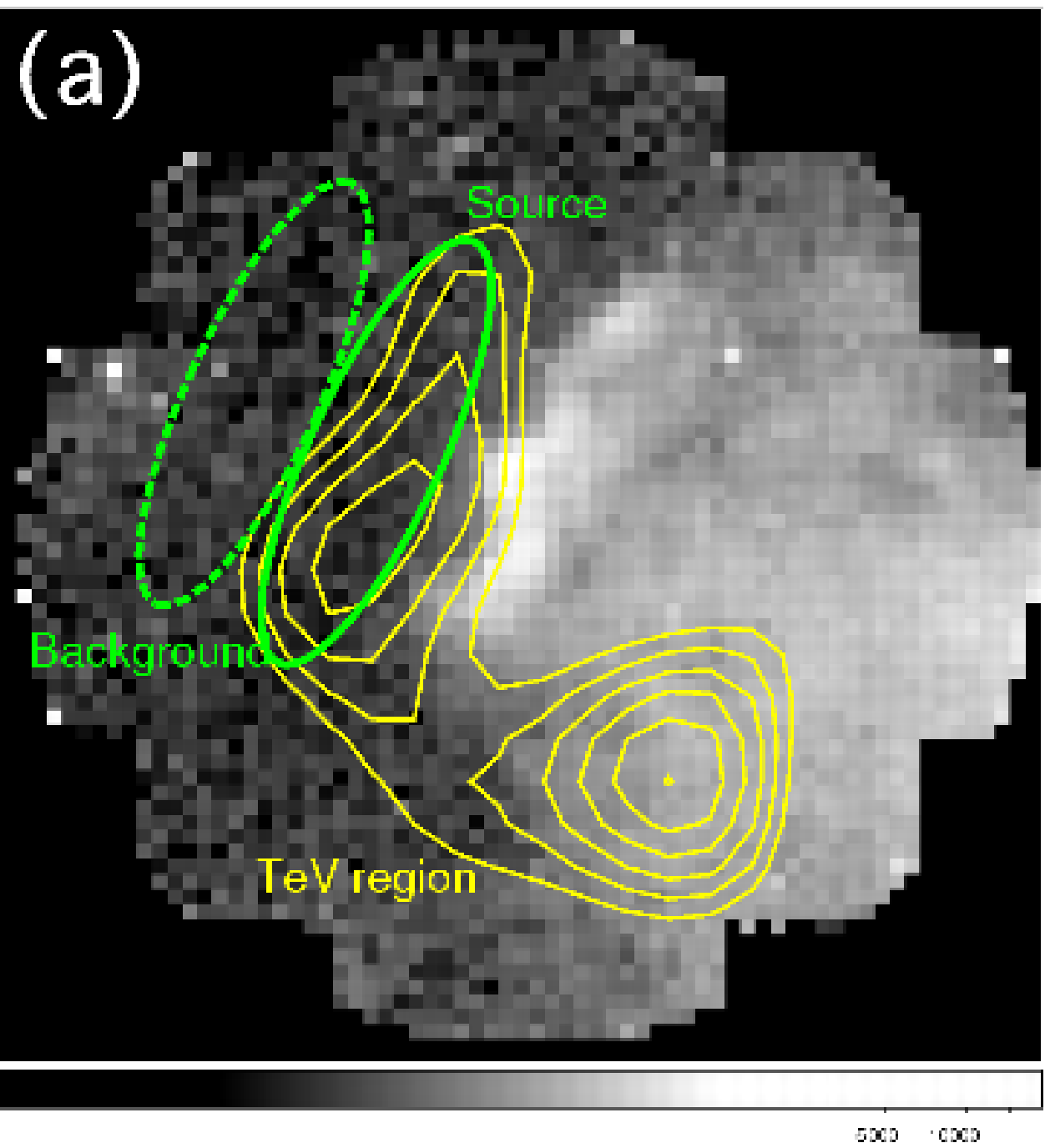}
    \includegraphics[width=50mm]{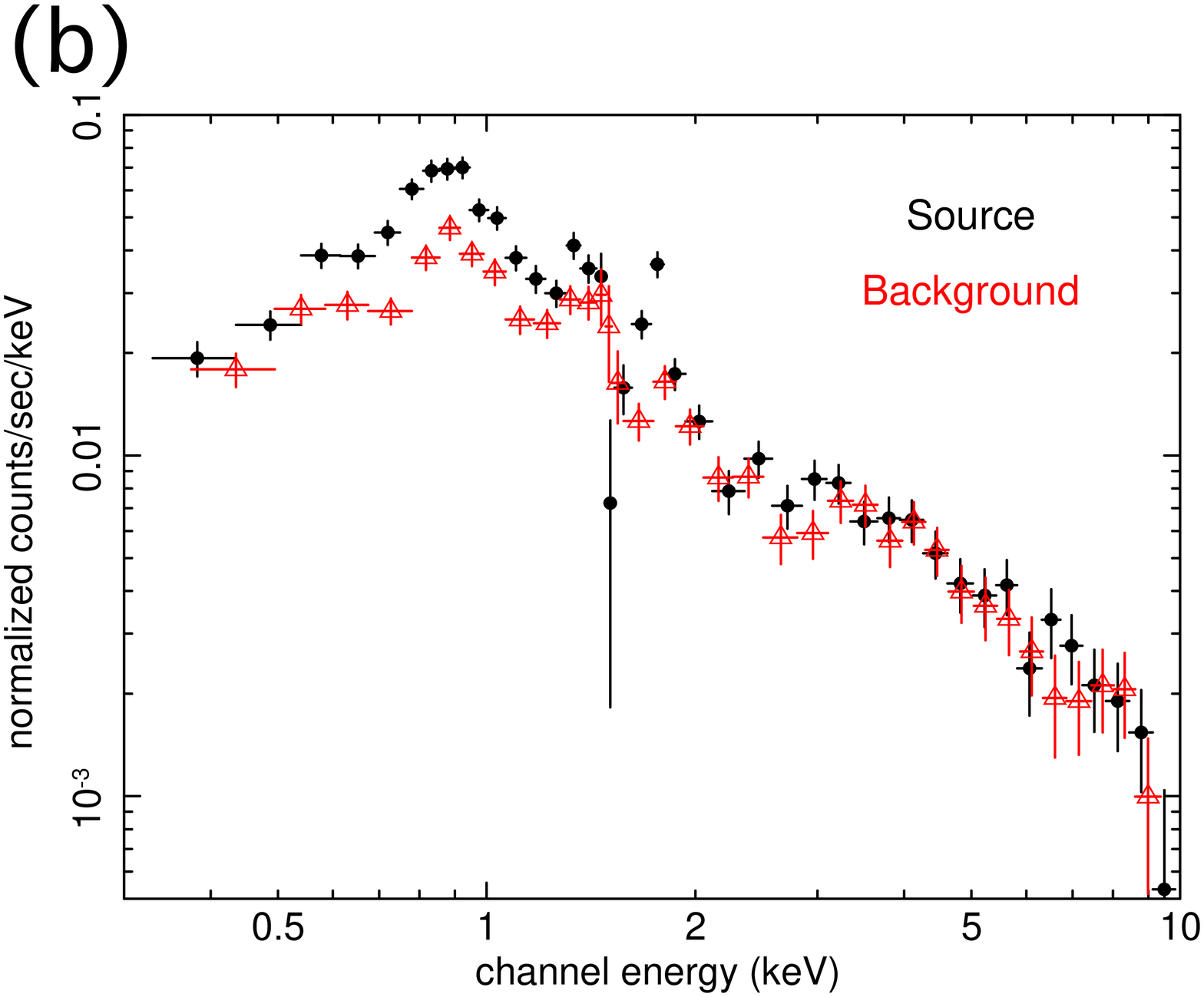}
    \includegraphics[width=55mm]{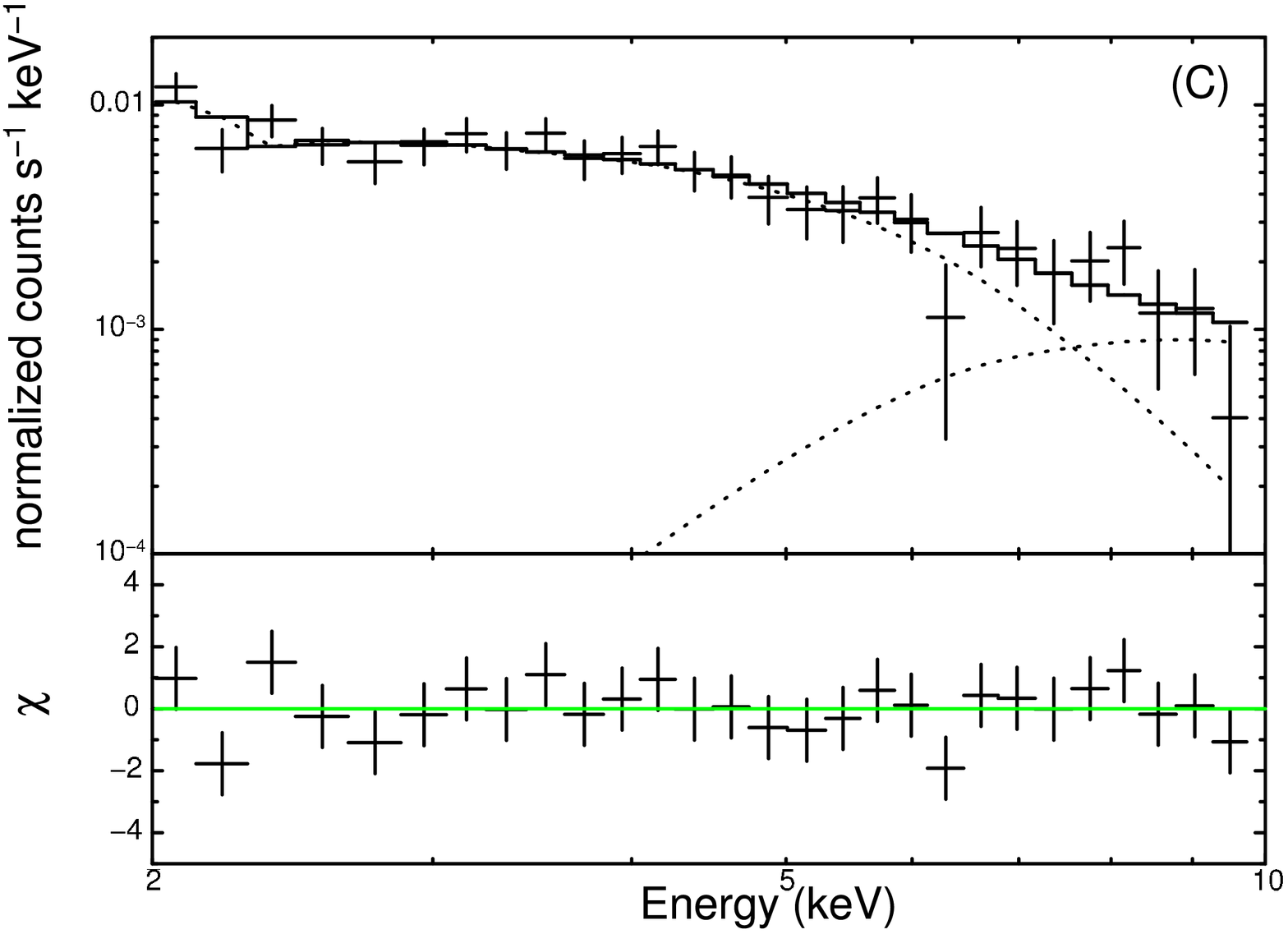}
  \end{center}
 \caption{Integration regions and resultant spectra of the north-eastern
 TeV $\gamma$-ray peak. (a) The source and background integration
 regions are drawn in green. They are an ellipse with a semi major and
 minor axis of $\timeform{6.'6}$ and $\timeform{1.'9}$,
 respectively. (b) The source (black) and background (red) spectra with
 the MOS instruments. NXB spectrum extracted from out of field of view
 is already subtracted. The source spectrum below 2~keV is contaminated
 by the thermal emission from the north-eastern rim.
(c) MOS background spectrum with the best-fit model.
}\label{fig:TeV_reg}
\end{figure}

We investigated X-ray spectrum of a region where intense TeV
$\gamma$--ray emission is detected in spite of no apparent X-ray
emission. In figure~\ref{fig:TeV_reg}(a) shown are the TeV $\gamma$-ray
intensity contours (the same as in figure~\ref{fig:allimage},
\cite{2008A&A...481..401A}) overlaid on the X-ray gray-scale image.
Although the TeV $\gamma$-ray surface brightness has a dual peak
structure,
% The yellow contour of figure \ref{fig:TeV_reg}(a) shows the count map of
% TeV $\gamma$--ray from \citet{2008A&A...481..401A}.
we concentrated on the north-eastern peak, and extracted the source
spectrum from the ellipse region located at ($l$, $b$) $\simeq$
($\timeform{6D.75}$, $-\timeform{0D.30}$), as shown in
figure~\ref{fig:TeV_reg}(a) in green, because the southern peak centered
at ($l$, $b$) $\simeq$ ($\timeform{6D.56}$, $-\timeform{0D.26}$) is
partly included in the south-eastern rim
and difficult to estimate nonthermal emission.
%which region is found
%to be contaminated by the thermal emission from W28.
The choice of the
source integration region adopted here also intends to exclude the
thermal emission from the north-eastern shell region.
%We note that this source region was used as the background region for the south-eastern shell analysis~(\S4.2).
The dashed ellipse in Figure~\ref{fig:TeV_reg} is,
on the other hand, a background region.
%The size with a major and minor axis radius is $\timeform{6.'6} \times \timeform{1.'9}$ for source and background regions. 
Figure~\ref{fig:TeV_reg}(b) shows the source and the background spectra thus
extracted, from which the Non X-ray Background~(NXB) spectrum evaluated
out of the telescope field of view are already subtracted. 
%Even we took
%care of the intrusion of the thermal emission from the north-eastern rim
%region, we found significant excess in the source spectrum below
%2~keV.
We have excess below 2~keV,
which may  be be stray light from W28 shell,
which is very difficult to estimate correctly.
Since we are interested in non-thermal emission associated with
the TeV $\gamma$-ray emission, we hereafter use the data only in the
band 2--10~keV.

The effective area of the background region is smaller than the source
region due to telescope vignetting. We checked the vignetting effect
by using the Lockman Hole archival data observed on 2002 December
27 to 29 (Obs~ID = 0147511601) with the total effective exposure time of
110~ks. A count ratio of the background to source regions at the same
detector positions was found to be 0.88 in the 2--10~keV band.
In evaluating flux of the possible non-thermal X-ray emission,
we first made an arbitral model of the background spectrum
to reconstruct the background emission in the source region.
Note that it is a mixture of Galactic ridge emission,
cosmic X-ray background, and position-dependent NXB,
so the resultant parameters have no physical meaning.
The two power-law model showed accepted fit
with the photon indices of 0.85 and $-$5.8,
as shown in Fig.~\ref{fig:TeV_reg}(c).
%we first
%tried to fit the background spectrum with a two power-law model,
%and
%found photon indices of 0.85 and $-$5.8, respectively.
The source
spectrum was fitted with a model comprising of the two power-law models
with the normalizations being multiplied by the vignetting-correction
factor, and an additional power-law model representing the putative
non-thermal X-ray emission, with the frozen photon index of 2.66 which is the
same as that in the TeV $\gamma$-ray band
\citep{2008A&A...481..401A}. Consequently, we only obtained an upper
limit for the non-thermal X-ray emission.
The 90\% upper limit flux in the
2--10~keV band is obtained to be 
$2.1\times10^{-14}$~ergs~cm$^{-2}$~s$^{-1}$.

%\newpage

\section{Discussion}
\subsection{The nature of the low temperature thermal component}
%\subsection{The nature of the thermal component}

We found the $\sim$0.3~keV thermal emission from the north-eastern shell
of W28
and deteremined the plasma parameters.
Table~\ref{tab:thermal_param} summarizes the electron
density ($n_{\rm e}$), the number of electrons ($N_{\rm e}$), the mass,
and the thermal energy ($E_{\rm thermal}$), on the basis of the best-fit
parameters summarized in table~\ref{tab:param_shell}.
\begin{table*}
\begin{center}
\caption{Thermal parameters}
\label{tab:thermal_param}
 \begin{tabular}{lcccccc}
 \hline \hline
 \multicolumn{1}{c}{Regions} & $V$ [cm$^{3}$] \footnotemark[a] & $n_{\rm e}$ [cm$^{-3}$] \footnotemark[b] & $n_{\rm e}t$ [10$^{11}$cm$^{-3}$s] \footnotemark[c] & $N_{\rm e}$ [10$^{56}$] & Mass [$M_\odot$] & $E_{\rm thermal}$ [10$^{48}$erg] \\
 \hline
% Shell region & & & &\\
  \hspace{1pc}1 & $1.3\times 10^{55}$ & 4.2$^{+0.7}_{-0.6}$ & 98~($\textgreater$4) & 0.54$^{+0.09}_{-0.08}$ & 0.044$^{+0.008}_{-0.007}$ & 0.090$^{+0.019}_{-0.015}$ \\
  \hspace{1pc}2 & $1.4\times 10^{55}$ & 9.4~$\pm0.9$ & 11~($\textgreater$6) & 1.3~$\pm0.1$ & 0.11~$\pm0.01$ & 0.18~$\pm0.02$ \\
  \hspace{1pc}3 & $1.7\times 10^{55}$ & 13~$\pm1$ & 50~($\textgreater$9) & 2.1~$\pm0.2$ & 0.17~$\pm0.01$ & 0.27~$\pm0.02$ \\
\hline 
Total & $4.4\times10^{55}$ & ---  & --- & 3.9$^{+0.4}_{-0.4}$ & 0.32$^{+0.03}_{-0.03}$ & 0.54~$\pm0.06$\\
\hline 
 \multicolumn{5}{@{}l@{}}{\hbox to 0pt{\parbox{160mm}{\footnotesize
 \footnotemark[a] Volumes were calculated on the assumption
 of an oval shape. See the text for
  more detail.\\
 \footnotemark[b] Electron density were obtained by using {\it E.M.} and
  the volume. \\
 \footnotemark[c] Ionization parameters~($n_{\rm e}t$) were found from
  the spectral fitting with a {\sc vnei} model as explained in \S~4.\\
 }\hss}}
 \end{tabular}
\end{center}
\end{table*}
We assume that the distance to W28 is 1.9~kpc
\citep{2002AJ....124.2145V}, and the plasma distributes uniformly along
the line of sight.
The shape of the emitting regions are assumed an oval
with the line-of-sight extent being the same as the semi-minor axis
appearing on the image ($r_{\rm s}$). The volumes are calculated to be
equal to $(4/3)\pi r_{\rm l} r^{2}_{\rm s}$, where $r_{\rm l}$ is the
length of the semi-major axis on the image.
% Calculated volumes~($V$) are summarized in table
% \ref{tab:thermal_param}. 
With the aid of the emission measure ($EM = \int n_{\rm e}n_{\rm H}dV$)
obtained from the spectral fitting and $V$, we calculated $n_{\rm e}$,
taking into account $n_{\rm e} \approx 1.24n_{\rm H}$ for fully ionized
solar abundance plasma. The number of electrons ($N_{\rm e} = n_{\rm
e}V$), the total mass and the thermal energy ($E_{\rm thermal} =
\frac{3}{2}(N_{\rm e}+N_{\rm H}+N_{\rm He})kT$) were obtained under the
assumption of energy equipartition between electrons and ions.  The
total thermal energy should be larger if other portions of the remnant
are included, and if the proton temperature is significantly larger than
the electron temperature as is expected for supernova remnants with
large shock velocities \citep{2007ApJ...654L..69G}.

\begin{figure}
  \begin{center}
    \includegraphics[width=70mm]{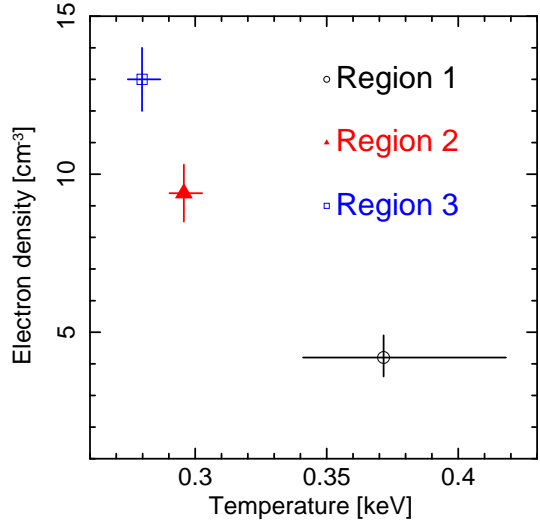}
  \end{center}
  \caption{(a)~Relation between the temperature and the electron density of shell regions.}
\label{fig:shell_kt_ne}
\end{figure}

The temperature and the electron density of the north-eastern shell
regions are plotted in figure~\ref{fig:shell_kt_ne}.
The value of $n_{\rm e}$ is much higher in the
north-eastern shell regions ($\sim$10~cm$^{-3}$) than
in the inner region ($\lesssim$1~cm$^{-3}: $\cite{2002ApJ...575..201R}).
In particular, shell region 3, which
apparently interacts with a molecular cloud (see figure~\ref{fig:ximage}
and \ref{fig:allimage}), has the largest $n_{\rm e}$.
% probably due to the compression by hitting the dense gas.
%
The higher temperature region has lower density. This is probably
because the original ejecta energy is distributed among larger amount of
interstellar matter for higher density regions, as indicated by the
total mass in the table.
% This suggests that the high electron density region is relatively
% getting cold by an adiabatic expansion. 
%The ionization parameters are, on the other hand, larger in the shell
%regions with $n_{\rm e}t\sim10^{13}$~s~cm$^{-3}$
%than in the inner region \citep{2012PASJ...64...81S}.
%This is the result of slow
%ionization due to low electron density in the inner region.
%The plasma
%age obtained with the combination of $n_{\rm e}t$ and $n_{\rm e}$ to be
%several times 10$^{4}$ year for both the shell and inner regions.
%The total volume of the region covered with the current study is about
%5\% of the whole SNR on the assumption that W28 is a sphere with a
%radius of 13~pc.  Given that the mass and the total energy in the
%north-eastern shell and inner regions are 5.9~$M_\odot$ and
%1.2$\times10^{49}$~ergs, respectively (table~\ref{tab:thermal_param}),
%the mass and the thermal energy of the entire SNR are estimated to be
%$\sim$120~$M_\odot$ and $\sim$2.4$\times10^{50}$~ergs, respectively. On
%the assumption that the electron density of the interstellar medium
%before a supernova expansion is 1~cm$^{-3}$, the total swept up mass by
%the shock is $\sim$200~$M_\odot$, which is roughly consistent with the mass
%estimate obtained from our spectral analysis.
% although there are the inaccuracy of the electron density of ISM, or
% the non-uniformity of emission from W28. Most X--ray seems to be
% emitted by swept up ISM by the shock  wave.
%

\subsection{Origin of the TeV $\gamma$-ray emission}

\begin{figure}[htpb]
  \begin{center}
    \includegraphics[width=80mm]{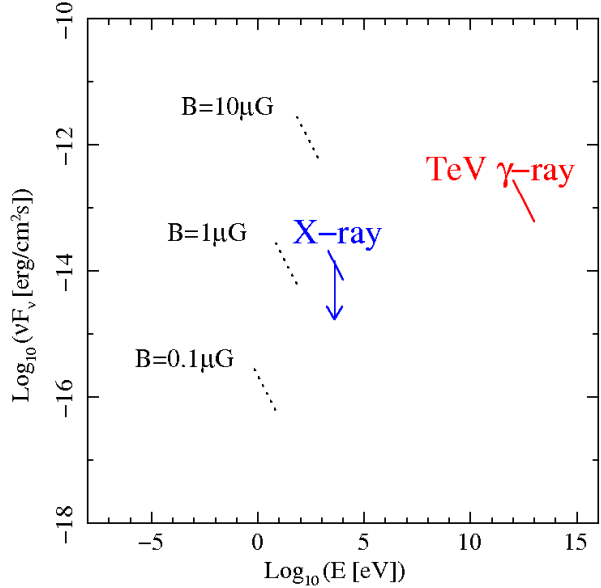}
  \end{center}
  \caption{Spectral energy distribution of TeV $\gamma$-ray region from
  the X-ray (blue) and TeV $\gamma$-ray (red). A photon index of the TeV
  $\gamma$-ray spectrum is observed to be 2.66, and the same index is
  assumed in drawing the X-ray upper limit. The dashed lines are the
  synchrotron radiation models with various magnetic
  field.}\label{fig:widespec}
\end{figure}

We obtained the upper limit of the X-ray flux from one of the TeV
$\gamma$-ray emission peak, as described in
\S~4.4. Figure~\ref{fig:widespec} shows the spectral energy distribution
of the TeV $\gamma$-ray and the X-ray flux upper limit in the region.
The red line is the TeV $\gamma$-ray power-law spectrum
\citep{2008A&A...481..401A} with a photon index of 2.66, while the blue
line is the upper limit of the X-ray power-law spectrum with a photon
index assumed to be the same as that in the TeV band. The TeV flux is
normalized to that of the X-ray flux by taking the difference in the
integration region into account (the X-ray region is the
$\timeform{6.'6}$ $\times$ $\timeform{1.'9}$ ellipse while the TeV
$\gamma$-ray one in \citet{2008A&A...481..401A} is a circle with a
radius of $\timeform{11.'8}$). The corrected TeV $\gamma$-ray flux is
3.3$\times10^{-13}$ ergs cm$^{-2}$ s$^{-1}$ in the 1--10~TeV band.

The flux ratio between 1--10~TeV and 2--10~keV
($F_{TeV}/F_X$) is often used
to examine the emission origin
\citep{2006MNRAS.371.1975Y}.
In young SNR and pulsar wind nebula (PWN) cases, the value is
in the range of $\sim 10^{-3}$--$1$
\citep{2007PASJ...59S.199M,2007PASJ...59S.209B},
whereas the value is larger than 16 in our case.
This result implies that
the $\gamma$-ray emission mechanism is different between
young SNRs/PWNe and our case of an old supernova remnant. 
Another sample is HESS~J1745$-$303,
which has the $F_{TeV}/F_X$ of larger than 4
\citep{2009ApJ...691.1854B}.
The nature of this source is still unknown,
but there is a molecular cloud and an old SNR G359.1$-$0.5
\citep{2000PASJ...52..259B,2011PASJ...63..527O}.

Such a high value of $F_{TeV/F_X}$ is similar to that of sources called
"dark particle accelerators"
which are $\gamma$-ray sources without any X-ray counterparts
\citep{2007PASJ...59S.199M,2007PASJ...59S.209B}.
Although their nature is still unknown,
one possibility can be old SNRs interacting with molecular clouds
like W28 and HESS~J1745$-$303.

Here, we consider the emission origin more precisely
using wide band spectrum.
We first
assume that the TeV $\gamma$-ray emission is powered through 1-zone
Inverse Compton scattering (IC) of the cosmic microwave background off
the accelerated primary electrons, and the same electrons emit
synchrotron radiation in the X-ray band. The characteristic energies of
a synchrotron photon $\epsilon$ and of an IC photon $E$ produced by the
same electron are related as
\begin{eqnarray}
\epsilon &\simeq& 0.07\;\left(\frac{E}{1\;{\rm TeV}}\right) \left(\frac{B}{10\;{\rm \mu G}}\right) \;\;\rm{[keV]}\\
%\;(B/10^{-5}\;{\rm G}) \;\;\rm{[keV]}\\
% \nonumber \\
f_{X} &\simeq& 10\;f_{TeV}\;\left(\frac{B}{10\;{\rm \mu G}}\right)^{2} 
% \nonumber
\end{eqnarray} 
where $B$ is the magnetic field and $f$ is the flux
\citep{aharonian_1997}. We draw synchrotron X-ray spectra with various
magnetic field ($B$ = 0.1, 1.0, and 10.0~$\mu$G) in
figure~\ref{fig:widespec} with dashed lines. We only obtained the upper
limit of the magnetic field of $\le$5~$\mu$G. This might be a little too weak for efficient particle acceleration region
\citep[for example]{2003ApJ...584..758V,2005ApJ...621..793B},
although it is possible that electrons accelerated in the past escape from the acceleration region and emit IC photons in the observed position where the magnetic field is weak. However, the TeV $\gamma$-ray
emission region well coincides with the distribution of the molecular
cloud, as demonstrated in \S~3 (figure~\ref{fig:allimage}). These facts
suggest that the TeV $\gamma$-ray emission originates from the pion
decay triggered by impacts of the high energy protons to the molecular
clouds.

\begin{figure}
  \begin{center}
\includegraphics[width=80mm]{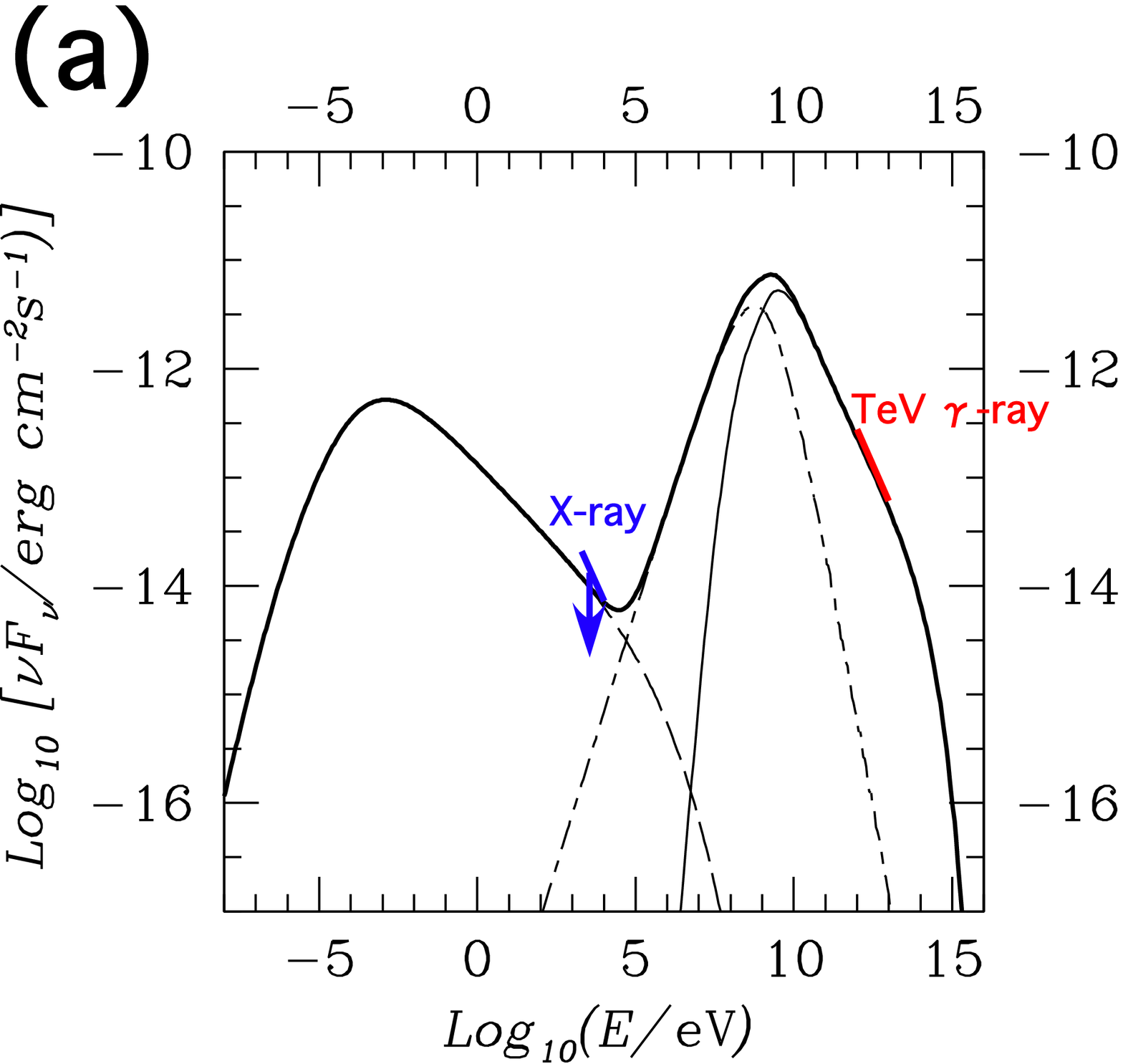}
\includegraphics[width=80mm]{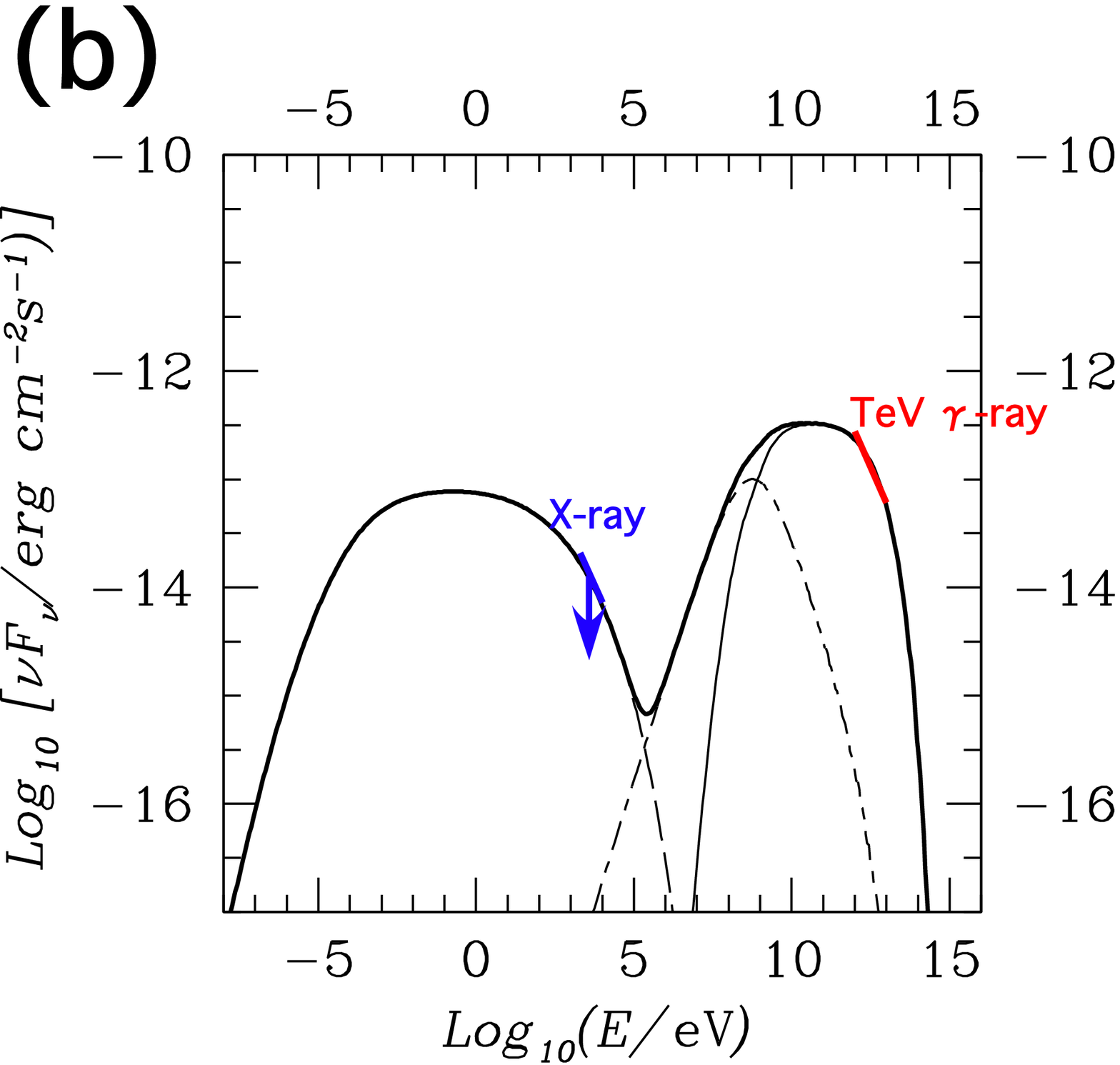}
  \end{center}
\caption{Spectra of hadronic emissions with the age, the distance and
the number density of molecular cloud of $t_{\rm age}$=3000~yr,
$D$=1.9~kpc and $n_{\rm mc}$=1000~cm$^{-3}$, respectively. Emissions are
$\pi^{0}$-decay $\gamma$-ray~(solid line), bremsstrahlung~(dot-dashed
line) and synchrotron~(dashed line) emission from secondary electrons
produced by charged pions. Total emissions are shown in bold line. Blue
and red line show upper limit X-ray and TeV $\gamma$-ray
spectra. (a)~With proton number index $p$=2.66, maximum proton energy
$E_{\rm max,p}$=1000~TeV, total energy $E_{\rm total}$=3.0 $\times
10^{47}$~erg, and magnetic field $B \lesssim$ 800 $\mu$G. (b)~With
$p$=2.0, $E_{\rm max,p}$=40~TeV, $E_{\rm total}$=8.1 $\times
10^{45}$~erg, and $B \lesssim$ 1500 $\mu$G.}  \label{fig:widespec_had}
\end{figure}

We then calculate a spectral energy distribution from X-ray to
TeV $\gamma$-ray bands based on the pion decay. In this calculation, the TeV
$\gamma$-ray originates from the process $\pi^0 \rightarrow \gamma +
\gamma$, where $\pi^0$ is produced through collisions between an
accelerated proton and a proton in a molecular cloud. Accordingly, the
resultant $\gamma$-ray spectrum depends on an energy distribution
function of the accelerated protons
\begin{equation}
 \frac{dN}{dE} \;=\; C\,E^{-p}\,{\rm exp}\left(-\,\frac{E} {E_{\rm max,p}} \right)\;\;(E_{\rm min,\,p} < E < E_{\rm max,\,p}),
\end{equation}
and the density of the molecular cloud,
where the energy $E$ in this context means the total energy.
%Of them, the minimum proton
%energy is taken to be equal to the half of the $\pi^0$ rest mass
%($E_{\rm min,\,p}$= 70 MeV). 
Here we set the minimum energy as the proton rest mass ($E_{\rm min,\,p}$= $m_{\rm p}\,c^{2}$). It is well known that the resultant
$\gamma$-ray energy is proportional to that of the bombarding proton
\citep{1994A&A...287..959D}, and hence we can simply adopt the number
index $p$ be equal to the photon index of the observed TeV $\gamma$-ray
spectrum (= 2.66). We assume the distance to W28 to be 1.9~kpc
\citep{2002AJ....124.2145V} to constrain the normalization $C$. The
density of the molecular cloud is measured from the CO observation
\citep{1999PASJ...51L...7A}, which is $10^3$ cm$^{-3}$. With these
parameters, we fitted the observed TeV $\gamma$-ray spectrum, and found
that $E_{\rm max,\,p} \gtrsim 10^{15}$~eV. The total energy of the accelerated
protons is obtained to be $3.0\times 10^{47}$ erg, which is much smaller
than the supernova explosion energy ($\simeq 10^{51}$ erg).
We have assumed a continuous and a homogeneous distribution of cosmic rays
in time and space, respectively, in the molecular cloud.
The result
is displayed in figure~\ref{fig:widespec_had}(a).
% Thus, we calculated 2 pattern spectra of TeV $\gamma$-ray with hadronic
% origin shown in figure~\ref{fig:widespec_had}.
%
Note that non-thermal bremsstrahlung originating from secondary
electrons produced through collisions between the accelerated protons
and the molecular cloud neutrons contribute significantly to the
resultant spectrum, although it dominates somewhat lower energy.
Actually, \citet{2010ApJ...718..348A} discussed that
the bremmsstrahlung-dominated emission model
requires unrealistic low magnetic field.
Given
$E_{\rm max,\,p}$ and the energy distribution of the secondary
electrons, which can be calculated from that of the protons, and the
absence of the synchrotron X-ray emission, we can set the upper limit on
the magnetic field strength. By referring to the upper limit of the
X-ray flux, we have obtained $B \lesssim$800 $\mu$G.
We remark that this upper limit of the
magnetic field was obtained by setting $E_{\rm max,
\,p} =$ 10$^{15}$~eV. In the case of $E_{\rm max, \,p} \, \textgreater$
10$^{15}$~eV, the model synchrotron spectrum
is enhanced and extends to higher energy without
being cutoff in the X-ray band.
Consequently, a weaker
magnetic field is required for the model synchrotron
X-ray spectrum to be accommodated with the X-ray flux upper limit. The
upper limit of $B =$800 $\mu$G is therefore the most conservative one in
the case of $p$=2.66.  The $B =$800 $\mu$G case is shown in
figure~\ref{fig:widespec_had}(a). The obtained parameters are summarized
in table~\ref{tab:hadron_param}.
\begin{table}[htb]
\begin{center}
\caption{Physical parameters obtained with hadronic model fitting}
\label{tab:hadron_param}
 \begin{tabular}{lcc}
 \hline \hline
 \multicolumn{1}{c}{Proton number index~($p$)} & (a)~2.66 & (b)~2.0 \\
 \hline
Maximum proton energy~($E_{\rm max,p}$) \ [TeV] & $\gtrsim$~1000 & 40\\
Total energy~($E_{\rm total}$) \ [erg] & 3.0 $\times$ 10$^{47}$ & 8.1 $\times$ 10$^{45}$ \\
Magnetic field~($B$) \ [$\mu$G] & $\lesssim$~800 & $\lesssim$~1500 \\
Photon flux in the 0.1--1~GeV \ [ph/cm$^{2}$s] &  2.3 $\times 10^{-8}$ & 6.3 $\times 10^{-10}$ \\
\hspace{18mm} in the 1--100~GeV \ [ph/cm$^{2}$s] &  4.1 $\times 10^{-9}$ & 1.6 $\times 10^{-10}$ \\
\hline \hline
% \multicolumn{3}{@{}l@{}}{\hbox to 0pt{\parbox{160mm}{\footnotesize
%     \footnotemark[a] \\
%     }\hss}}
 \end{tabular}
\end{center}
\end{table}

We note that, at the old SNR shocks with low shock velocity,
the maximum energy of accelerated particles is smaller due to the escape of high-energy protons~(e.g. \cite{2010A&A...513A..17O}).
 We thus have tried to calculate the spectrum in the case that the
number index of the proton is equal to the canonical value of the Fermi
acceleration ($p = 2.0$) in the lower-energy side. In this case, we adjust
$E_{\rm max,\,p}$ so that the resultant model spectrum can be fitted to
the observed TeV $\gamma$-ray spectrum. The result of the maximum-$B$
case is shown in figure~\ref{fig:widespec_had}(b). The maximum proton
energy is found to be $E_{\rm max,\,p} = 40$~TeV, and the total proton
energy is $8.1\times 10^{45}$~erg, which is again much smaller than the
supernova explosion energy. The upper limit of the magnetic field
strength is $B \lesssim$1500$\mu$G.

%\citet{2009ApJS..183...46A} reported
%based on the Fermi observation that the
%integrated photon fluxes in the
%bands 0.1--1~GeV and 1--100~GeV are
%$7.42\times 10^{-7}$ photons~cm$^{-2}$~s$^{-1}$ and
%$4.51\times 10^{-8}$ photons~cm$^{-2}$~s$^{-1}$,
%respectively.  
%which are equal to
%$\sim$~1.0~$\times~10^{-9}$~erg/cm$^{2}$s and
%$\sim$~1.0~$\times~10^{-10}$~erg/cm$^{2}$s with our model,
%respectively.
%On the other hand, the photon fluxes in the same energy bands in our
%models are calculated to be $2.0\times 10^{-8}$
%ph~cm$^{-2}$~s$^{-1}$ and $2.7\times 10^{-9}$
%ph~cm$^{-2}$~s$^{-1}$, respectively, for the case of
%figure~\ref{fig:widespec_had}(a), and $5.6\times 10^{-10}$
%ph~cm$^{-2}$~s$^{-1}$ and $2.0\times 10^{-10}$
%ph~cm$^{-2}$~s$^{-1}$, respectively, for the case of
%figure~\ref{fig:widespec_had}(b), shown in table~\ref{tab:hadron_param}, both of which are smaller than the
%values obtained by Fermi. This is because the
%Although our calculated flux in the GeV band of $\lesssim
%10^{-11}$~erg/cm$^{2}$s~(see figure~\ref{fig:widespec_had}) 
%source integrated region of {\it Fermi} is larger than and slightly
%different from our integrated region
%(figure~\ref{fig:TeV_reg}a). Hence, the flux of our
%models can be regarded as being consistent with
%that of the GeV emission.
%More detailed observations will be needed
%to construct wide-band spectra
%including GeV data in smaller scale.

\section{Conclusion}

We analyzed the XMM-Newton data of the north-eastern part of the
supernova remnant W28. The observed X-ray image showed
the bright and twisted north-eastern shell
and the inner emission region which is
part of the center-filled emission brightening toward the south-west end
of the field of view.
% is the brightest, and hard emission was discovered from inner
% region. TeV $\gamma$-ray well coincide with the molecular cloud
% distribution.

The X-ray emission from north-eastern shell is found
to reach collisional ionization equilibrium
state and can be fitted well with a single temperature optically thin
thermal emission model with $kT$ of $\simeq$0.3~keV. From the emission
measure and the apparent volume, the electron density is found to be as
high as $\simeq$10~cm$^{-3}$. Since a bunch of molecular cloud spatially
coincides with the outer edge of a part of the shell, this high density
is due to the collision of the plasma with the molecular
cloud.
% caused this high electron density.

In contrast, there is no significant X-ray emission from one of the TeV
$\gamma$-ray peaks. We only obtained the 90\% upper limit flux of
2.1$\times$ 10$^{-14}$ erg cm$^{-2}$ s$^{-1}$ in the 2--10~keV band,
assuming a power-law spectrum with the same photon index as in the TeV
$\gamma$-ray (= 2.66). The spatial coincidence of the molecular cloud
and the TeV $\gamma$-ray emission site suggests that TeV $\gamma$-ray is
hadronic origin.
We calculate the spectra of hadrons including
pions, kaons, nucleons and so on 
produced through proton-proton scatterings and their daughter
particles (gamma-rays, electrons, neutrinos and so on),
 and found
that $\pi^{0}$-decay emission is dominant in the TeV $\gamma$-ray band. A
weak upper limit on the magnetic field strength is obtained $B\lesssim$
1500$\mu$G from the X-ray flux upper limit.

\section*{Acknowledgements}
We thank the anonymous referee for fruitful comments.
This work was supported in part by Grant-in-Aid for Scientific Research
of the Japanese Ministry of Education, Culture, Sports, Science and Technology,
No.~22684012 (A.~B.), No.~24840036 (M.~S.), No.~21111006, 22244030,
and 23540327 (K.K.).
Two of authors (A.~B. and R.~Y) would like to express their deep
appreciation for the Research Institute, Aoyama-Gakuin University, for
supporting our research.

\end{document}